# First-Principles Study of Two-Dimensional CeN: Interplay between Structural Stability, Electronic Structure, and Magnetism

Jose Mario Galicia Hernandez[1], Jose Jorge Rios Ramirez[2], Gregorio Hernandez Cocoletzi[1] and Juan Francisco Rivas Silva[1].

[1] *Instituto de Física "Ing. Luis Rivera Terrazas", Benemérita Universidad Autónoma de Puebla. Av. San Claudio & Blvd. 18 Sur, Puebla, Mexico, 72570.*

[2] *Facultad de Ingeniería, Benemérita Universidad Autónoma de Puebla. Blvd. Valsequillo & Av. San Claudio, Puebla, Mexico, 72570.*

***Abstract:*** In this work, the structural stability, and the electronic and magnetic properties of a novel CeN monolayer derived from the (111) surface of bulk rock salt CeN are investigated using first-principles calculations. Dimensional reduction produces substantial changes in the local atomic environment, decreasing the Ce–N coordination from sixfold in the bulk to threefold in the monolayer and shortening the Ce–N bond length, while preserving an essentially planar hexagonal structure. Moreover, the electronic and magnetic properties are significantly modified, which may be useful for practical applications. Both ferromagnetic (FM) and antiferromagnetic (AFM) configurations were investigated and found to be strongly bound, with the FM state slightly lower in energy. The magnetic moments are predominantly localized on the Ce atoms and mainly originated from the Ce 4*f* states. The electronic structure also exhibits a pronounced dependence on magnetic ordering, highlighting the connection between localized 4*f* states and the spin-dependent electronic properties of the system. In addition, the calculated elastic properties confirm the mechanical stability of the monolayer, while phonon calculations support its dynamical stability. Ab initio molecular dynamics simulations further indicate that the two-dimensional structure preserves its integrity under finite-temperature conditions. Overall, these results demonstrate that dimensional reduction significantly modifies the structural, electronic, and magnetic behavior of CeN without compromising the stability of the resulting two-dimensional lattice. The combination of structural robustness, localized Ce 4*f* magnetism, and spin-dependent electronic properties make the CeN monolayer an interesting system for further exploration in two-dimensional magnetism and spin-dependent applications.

## 1. Introduction

Since the discovery of graphene, the design and synthesis of atomically thin materials have shown a growth in the fields of nanosciences, nanotechnology, material science and condensed matter physics, allowing the fabrication of several devices with novel and outstanding electronic, magnetic, optical, and mechanical properties **[1, 2]**. In recent years, the work on 2D systems has not been limited to graphene; a broad range of materials have been studied, including transition-metal dichalcogenides, phosphorene, MXenes, among others. All these materials have outstanding properties allowing potential applications in nanoelectronics, optoelectronics, spintronics, as well as quantum technologies **[2-5]**. These properties can be explained in terms of the quantum confinement effects because of the reduction in dimensionality, leading to modifications in electronic interactions and alterations in the magnetic exchange mechanisms not observed on bulk counterparts. On the other hand, the search for materials with intrinsic magnetism has attracted considerable attention, especially those where the magnetic phenomena can survive even on the monolayer limit **[3, 4]**. This fact gives rise to extensive efforts aim to develop low-dimensional magnetic materials with strong spin polarization suitable for practical applications, especially on spintronics **[5]**. With this in mind, it is not surprising that compounds containing rare-earth elements are good candidates for these objectives, because of the presence of partially 4*f*-orbitals, which are responsible of large magnetic moments and high correlation effects. Consequently, rare-earth materials are part of a very important class of strong correlated systems in which an interplay between localization and itinerant behavior of electrons is observed, giving rise to their particular electronic and magnetic properties, such as highly localized magnetic moments and itinerant electronic states.

Among the rare-earth elements, cerium (Ce) possesses particular properties due to its single 4*f*-electron, which owns a localized and itinerant behavior, this fact makes the cerium-based materials to show electronic properties strongly influenced by the competition between electron localization, crystal field and hybridization among Ce 4*f* states and neighbor orbitals **[6-9]**. As a result, cerium compounds have been widely studied by using several theoretical frameworks, for understanding its highly correlated effects **[9-12]**. Cerium mononitride (CeN) belongs to the family of rare-earth nitrides, a special kind of compounds that has recently attracted significant attention because of its electronic, magnetic and spin-dependent transport properties **[13-15]**. Both experimental and theoretical studies have been performed to demonstrate that this compound exhibits half-metallic and highly spin-polarized electronic structure, suggesting that this material is an attractive candidate for building spin-filtering and spin-injection devices **[13-15]**. In comparison with other rare-earth nitrides, CeN has unusual electronic behavior because of the nature of its Ce 4*f* electrons; they hybridize with Ce 5*d* and N 2*p* orbitals, resulting in an itinerant and mixed-valence electronic character **[9-12]**. This behavior differs from that observed in other mononitrides, where the 4-*f* electrons are commonly described as localized, as the states of these electrons are situated far from Fermi level and primarily contribute to stabilize the magnetic moments. It has been reported that CeN shows a 4*f*-band formation, suggesting that the compound is located close to the itinerant side of the localization-delocalization crossover **[10]**. Other works confirmed that CeN exhibits a pronounced Ce 4*f*-N 2*p* hybridization and significant valence fluctuations **[11, 12]**.

On the other hand, the study of a novel 2D CeN is motivated by the possibility of discovering new magnetic functionalities. The coexistence of large magnetic moments, electronic correlations and reduction of dimensionality, provides a favorable environment for the emergence of unconventional magnetic states **[15-17]**. In recent theoretical studies it has been shown that low-dimensional rare-earth compounds can exhibit robust magnetic ordering and high spin-polarized electronic structures, which are desirable properties for spintronic applications **[16-17]**. In this way, CeN is a unique compound to be investigated in the competition between electron localization, hybridization, and magnetic interactions in a correlated *f*-electron system. Understanding how these competing mechanisms evolve under the reduction of dimensionality is of fundamental interest, as dimensional confinement may significantly modify the hybridization strength, magnetic interactions and orbital occupation that do not exist in the bulk phase. Despite the extensive literature devoted to bulk CeN, the study of properties of 2D-CeN remains still largely unexplored, especially the structural stability, electronic band structure, the orbital-resolved density of states and the magnetic behavior. Therefore, 2D-CeN offers an excellent opportunity to explore how dimensional reduction affects the behavior of correlated 4*f* electrons, and how these effects influence the resulting electronic and magnetic properties.

In this paper, we performed first-principles calculations to investigate the structural, electronic and magnetic properties of 2D-CeN. Particular attention is devoted to study the role of Ce 4*f*-states, to explain the nature of the magnetic ordering, the orbital hybridization and to describe the electronic states in the vicinity of Fermi level. With this study, we expect to elucidate the interplay among reduced dimensionality, electronic correlations, and magnetism, to provide valuable insights into the physics of low-dimensional rare-earth nitrides. This paper is organized as follows, in section 2 we describe in detail the computational methods used for calculations. The results and discussions about structural and electronic properties are presented in section 3. Finally, a summary of results and conclusions are presented in section 4.

## 2. Computational methods

Computations were performed by using the periodic Density Functional Theory as implemented in Vienna Ab initio Simulation Package (VASP) **[18, 19]**. The exchange-correlation energy was treated within the Generalized Gradient Approximation (GGA) in the parametrization of Perdew-Burke-Ernzerhof (PBE) **[20]**. The electron-ion interaction was modeled employing PAW pseudopotentials **[21, 22]**. The first Brillouin zone was sampled by using a regular k-point mesh in the parametrization of Monkhorst-Pack **[23]**. For ground state calculations, we used a mesh of $9\times9\times9$ for the bulk system and $9\times9\times1$ for monolayers. For computing the electronic properties, the mesh used was $19\times19\times19$ and $19\times19\times1$ for the bulk and monolayers respectively. The wave function was expanded using plane waves with a kinetic energy of 600 eV. For the structure optimization, we used conjugated gradient algorithm. The ionic relaxation was achieved when the forces on each atom were less than 0.01 eV/Å. On the other hand, the convergence criteria for the energy difference between two steps in the self-consistent cycle was set to $10^{-8}$ eV. The layers were modeled by using the supercell method, for which a vacuum space of 15 Å was set in the *c*-direction to avoid the interaction between adjacent slabs. As the system includes Ce atoms, Hubbard correction **[24]** was necessary for treating the electronic correlation effects, in this way, the correction was included in the parametrization of Dudarev et al. **[25]**, the on-site Coulomb interaction was set to 4.5 eV for Ce-*f* orbitals in all systems.

## 3. Results and discussions

The results are presented and discussed in this section by systematically examining the structural, energetic, magnetic, electronic, mechanical, dynamical, and thermal properties of two-dimensional CeN, with particular emphasis on the changes induced by dimensional reduction with respect to the parent bulk phase.

*3.1 Structural properties*

The structural properties of bulk and two-dimensional CeN were investigated after full structural relaxation. In the first step, we relaxed the bulk structure. We used the conventional cubic cell, belonging to the rocksalt (or halite) structure, with space group $Fm\bar{3}m$ (No. 225) consisting of 4-Ce atoms and 4-N atoms. The two-dimensional system was constructed starting from a single layer derived from the (111) orientation of bulk CeN. This construction provides a direct structural connection between the three-dimensional parent compound and the investigated two-dimensional phase and allows the effects of dimensional reduction on the local Ce–N environment to be analyzed.

The optimized bulk CeN structure retains the cubic rocksalt geometry, with four Ce atoms and four N atoms in the conventional unit cell, corresponding to four CeN formula units. The optimized lattice parameter was found to be a = b = c = 5.291 Å, in good agreement with the reported values (5.02 Å **[26]**, 5.018 Å **[27]**, 5.024 Å **[28]**). In bulk structure, the resulting local environment is therefore characterized by sixfold octahedral coordination. For the optimized lattice constant, the nearest-neighbor Ce–N distance was found to be 2.645 Å. This structure provides a reference geometry from which the structural changes associated with dimensional reduction can be evaluated.

The two-dimensional CeN structure was generated from a single layer associated with the (111) surface of the rock-salt bulk phase. The resulting layer possesses a hexagonal in-plane geometry. A 2×2 supercell containing four Ce and four N atoms was employed for the calculations, together with a vacuum space of 15 Å along $c$-axis to avoid interactions between adjacent layers. A 2×2 supercell was constructed to properly investigate the physical properties of the AFM configuration, since the 1×1 unit cell does not provide sufficient degrees of freedom to accommodate the two sets of oppositely oriented magnetic moments required to represent the considered AFM ordering.

After structural relaxation, the computed lattice constants for the FM monolayer were equal to a = b = 8.096 Å. Since the calculations were performed using a 2×2 cell, the corresponding primitive lattice parameter is approximately equal to 4.048 Å.

The AFM structure exhibits an essentially identical equilibrium geometry, with a characteristic primitive lattice parameter of 4.050Å, and for the 2×2 supercell, lattice parameters equal to a = b = 8.100 Å. Therefore, both the FM and AFM monolayers are considered to preserve the hexagonal symmetry of the two-dimensional CeN structure.

The relationship between the bulk and monolayer lattice parameters can also be understood from the geometry of the rock-salt (111) orientation. For the optimized cubic bulk lattice constant, the characteristic in-plane periodicity associated with the (111) plane is 3.742 Å.

The relaxed monolayer values of lattice constants of 2D structures are therefore larger than the corresponding bulk-derived in-plane periodicity, indicating an appreciable in-plane relaxation upon isolation of the single layer. The relative increase is approximately 8.2%.

This expansion reflects the substantial reorganization of the atomic environment when the three-dimensional rock-salt coordination network is reduced to an isolated two-dimensional layer.
One of the most significant consequences of dimensional reduction is the change in the local atomic coordination. In bulk rock-salt CeN, every Ce atom has six nearest-neighbor N atoms, whereas every N atom is likewise coordinated by six Ce atoms. In contrast, in the 2D structures, each Ce atom has three nearest-neighbor N atoms, and each N atom has three nearest-neighbor Ce atoms, thus, dimensional reduction changes the local coordination. This reduction is accompanied by a significant relaxation of the remaining Ce–N bonds. For the FM monolayer, the nearest-neighbor distance is approximately 2.337 Å, whereas the AFM configuration gives a very similar average value of approximately 2.341 Å. Compared with the bulk value of 2.646 Å, the Ce–N bonds in the monolayer are therefore shortened by approximately 0.31Å. This corresponds to a relative contraction of around 11.7%. The shortening of the Ce–N bonds can be associated with the substantial reduction of the coordination environment. The remaining Ce–N interactions consequently undergo structural relaxation, leading to shorter equilibrium bond distances. The simultaneous reduction in coordination and contraction of the remaining Ce–N bonds therefore represent the principal local structural consequences of dimensional reduction in CeN.

Finally, structural relaxation does not produce an out-of-plane reconstruction in both 2D structures. Consequently, both magnetic configurations can be described as essentially planar CeN monolayers. The preservation of planarity is an important structural characteristic of the optimized system. Although extraction of the layer from the three-dimensional parent structure strongly modifies the atomic coordination and Ce–N bond distances, it does not induce an appreciable out-of-plane reconstruction. The principal relaxation therefore occurs within the plane of the monolayer. Another important fact to consider is that changing the magnetic configuration from FM to AFM produces only negligible structural modifications, so magnetic ordering has only a minor effect on the equilibrium geometry, suggesting that the structural framework of the CeN monolayer is largely insensitive to whether the Ce magnetic moments adopt FM or AFM ordering. The relaxed structures of bulk CeN and 2D CeN monolayers are depicted in figure 1.

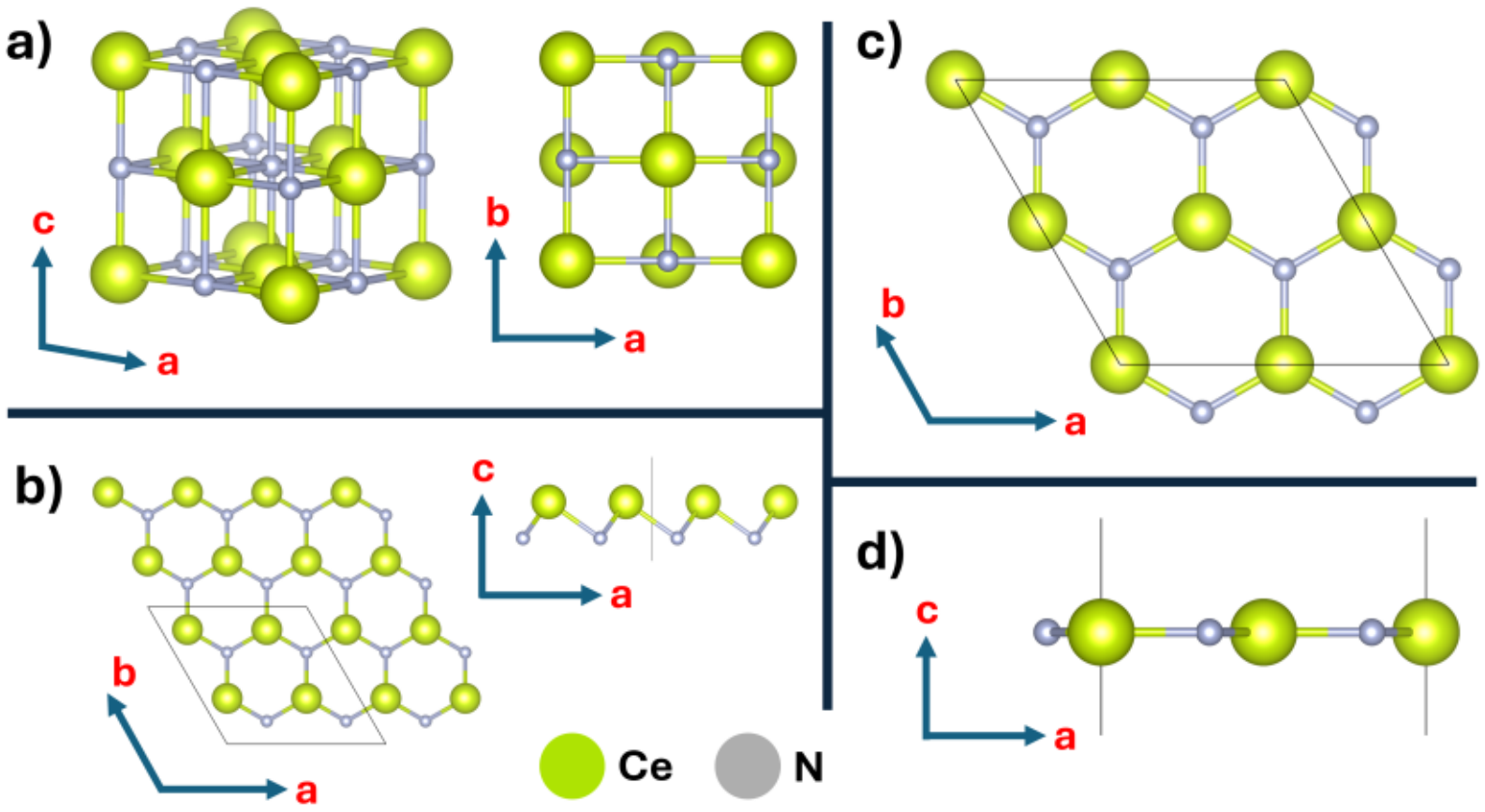

**Figure 1.** Relaxed structures of CeN, a) Side view and top view of bulk CeN, b) side view and top view of a single layer of the bulk (111) surface, c) top view of FM/AFM 2D CeN monolayer, and d) side view of FM/AFM 2D CeN monolayer showing its planar structure.

*3.2 Cohesive and formation energies*

The energetic stability of the CeN monolayer was evaluated through the cohesive and formation energies. Cohesive energy describes the energetic stabilization of the bonded structure relative to its constituent isolated atoms, whereas the formation energy measures the energetic favorability of CeN formation relative to the corresponding elemental reference states. Bulk CeN was also considered as a reference to assess how dimensional reduction affects the energetic stability of the material.

The cohesive energy per atom was calculated according to:

$$E_{\mathrm{coh}} = \frac{E_{\mathrm{sys.}} - n_{\mathrm{Ce}} E_{\mathrm{Ce}}^{\mathrm{atom}} - n_{\mathrm{N}} E_{\mathrm{N}}^{\mathrm{atom}}}{n_{\mathrm{Ce}} + n_{\mathrm{N}}} \qquad (1),$$

where $E_{\mathrm{coh}}$ is the cohesive energy per atom, $E_{\mathrm{sys.}}$ is the total energy of the system under consideration, $E_{\mathrm{Ce}}^{\mathrm{atom}}$ and $E_{\mathrm{N}}^{\mathrm{atom}}$ are the total energies of isolated Ce and N atoms, respectively, and $n_{\mathrm{Ce}}$ and $n_{\mathrm{N}}$ represent the numbers of Ce and N atoms in the system respectively.

The formation energy provides a complementary measure of stability by comparing the energy of CeN with those of its constituent elements in selected reference states. In the present case, elemental Ce bulk and molecular nitrogen were used as reference states. The formation energy per atom was calculated as:

$$E_{\mathrm{f}}^{\mathrm{Ce-N}} = \frac{E_{\mathrm{sys.}} - n_{\mathrm{Ce}} E_{\mathrm{Ce}}^{\mathrm{bulk}} - n_{\mathrm{N}} E_{\mathrm{N}_2}}{n_{\mathrm{Ce}} + n_{\mathrm{N}}} \qquad (2),$$

where $E_{\mathrm{f}}^{\mathrm{Ce-N}}$ is the formation energy per atom with respect to elemental Ce bulk and molecular nitrogen, $E_{\mathrm{sys.}}$ is the energy of the system under consideration, $E_{\mathrm{Ce}}^{\mathrm{bulk}}$ is the calculated energy per atom of elemental bulk Ce and $E_{\mathrm{N}_2}$ is the total energy per atom of an isolated nitrogen molecule, and $n_{\mathrm{Ce}}$ and $n_{\mathrm{N}}$ are the number of Ce and N atoms in the system respectively.

On the other hand, we can compute the formation energies of 2D structures with respect to FM bulk, which is the most stable form of CeN by using the following equation:

$$E_{\mathrm{f}}^{\mathrm{CeN}} = \frac{E_{\mathrm{sys.}} - E_{\mathrm{Ce-bulk}}^{\mathrm{FM}} - n_{\mathrm{Ce}} E_{\mathrm{Ce}}^{\mathrm{bulk}} - n_{\mathrm{N}} E_{\mathrm{N}_2}}{n_{\mathrm{Ce}} + n_{\mathrm{N}}} \qquad (3),$$

where $E_{\mathrm{f}}^{\mathrm{CeN}}$ is the formation energy with respect to FM bulk, $E_{\mathrm{sys.}}$ is the energy of the system under consideration, $E_{\mathrm{Ce-bulk}}^{\mathrm{FM}}$ is the total energy of bulk CeN (FM), taken as the reference energy, $E_{\mathrm{Ce}}^{\mathrm{bulk}}$ is the energy per atom of bulk Ce, $E_{\mathrm{N}_2}$ is the energy per atom of $N_2$ molecule, and $n_{\mathrm{Ce}}$ and $n_{\mathrm{N}}$ are the number of Ce and N atoms in the system respectively.

The calculated cohesive and formation energies are summarized in table 1.

**Table 1.** Cohesive and formation energies for bulk FM CeN and the 2D CeN structures.

| System | $E_{\mathrm{coh}}$ (eV/atom) | $E_{\mathrm{f}}^{\mathrm{CeN}}$ (eV/CeN) | $E_{f}^{\mathrm{Ce-N}}$ (eV/atom) |
|---|---|---|---|
| **Bulk FM** | −7.552 | ------- | −6.119 |
| **2D FM** | −6.693 | 2.760 | −5.260 |
| **2D AFM** | −6.691 | 2.762 | −5.258 |

In table 1, a negative sign in energy indicates that the system in consideration is more stable than the reference. On the other hand, a positive sign implies that the system under consideration is less stable than the reference. In this way, bulk CeN and both configurations of 2D CeN are more stable than the Ce and N isolated atoms and, furthermore more stable than their corresponding constitutive species: Ce in bulk phase and $N_2$ molecular nitrogen; more negative values of cohesive or formation energies indicate stronger bonds between atoms. On regards the stability of 2D systems with respect to bulk CeN, positives values of formation energies were obtained, from this, we can conclude that the bulk phase of CeN is more stable than the 2D counterparts, which is not surprising, as bulk systems are in general more stable than the 2D freestanding equivalent systems, thus this positives values are not related to instabilities of 2D structures, it is just an indication that bulk is more stable than 2D structures.

The bulk FM structure exhibits a more negative cohesive energy than the monolayer, the difference between bulk FM and 2D FM is approximately 0.859 eV/atom. This reduction in the magnitude of the cohesive energy is consistent with the change from the three-dimensional coordination network of bulk CeN to the reduced coordination environment of the monolayer. Nevertheless, the relatively large negative cohesive energy of the two-dimensional structure demonstrates that dimensional reduction does not eliminate the strong bonding between Ce and N. On the other hand, the FM and AFM monolayers exhibit almost identical cohesive energies. Their difference is only approximately 2.1 meV/atom. This small difference is consistent with the small energetic separation previously obtained between the FM and AFM magnetic configurations and indicates that both magnetic arrangements possess essentially the same degree of atomic cohesion.

The FM and AFM monolayers again exhibit almost identical values. Their formation-energy difference is approximately 2 meV/atom. This small difference reflects the close energetic competition between the two magnetic configurations. The FM state remains slightly more favorable, but the energetic stability of the monolayer with respect to its constituent elements is essentially unaffected by whether the Ce local moments adopt FM or AFM ordering.

In conclusion, taken together, the cohesive and formation energies provide complementary evidence for the energetic stability of the 2D CeN monolayers. The highly negative cohesive energies demonstrate substantial stabilization relative to isolated Ce and N atoms, indicating strong atomic cohesion within the two-dimensional structures. Similarly, the highly negative formation energies obtained relative to bulk Ce and molecular nitrogen indicate favorable formation within the elemental reference scheme considered.

*3.3 Electronic properties*

The electronic properties of the two-dimensional CeN system were investigated for both ferromagnetic (FM) and antiferromagnetic (AFM) configurations and compared with those of bulk CeN in order to elucidate the effect of dimensional reduction on its electronic and magnetic behavior.

In the first step, we computed the electronic band structure and projected density of state of bulk system for comparison purposes with 2D, for exploring the modification in electronic behavior because of reduction in dimensionality. The calculations were done for FM phase, which is reported as the most stable for bulk. Figure 2 depicts the electronic band structure and projected DOS.

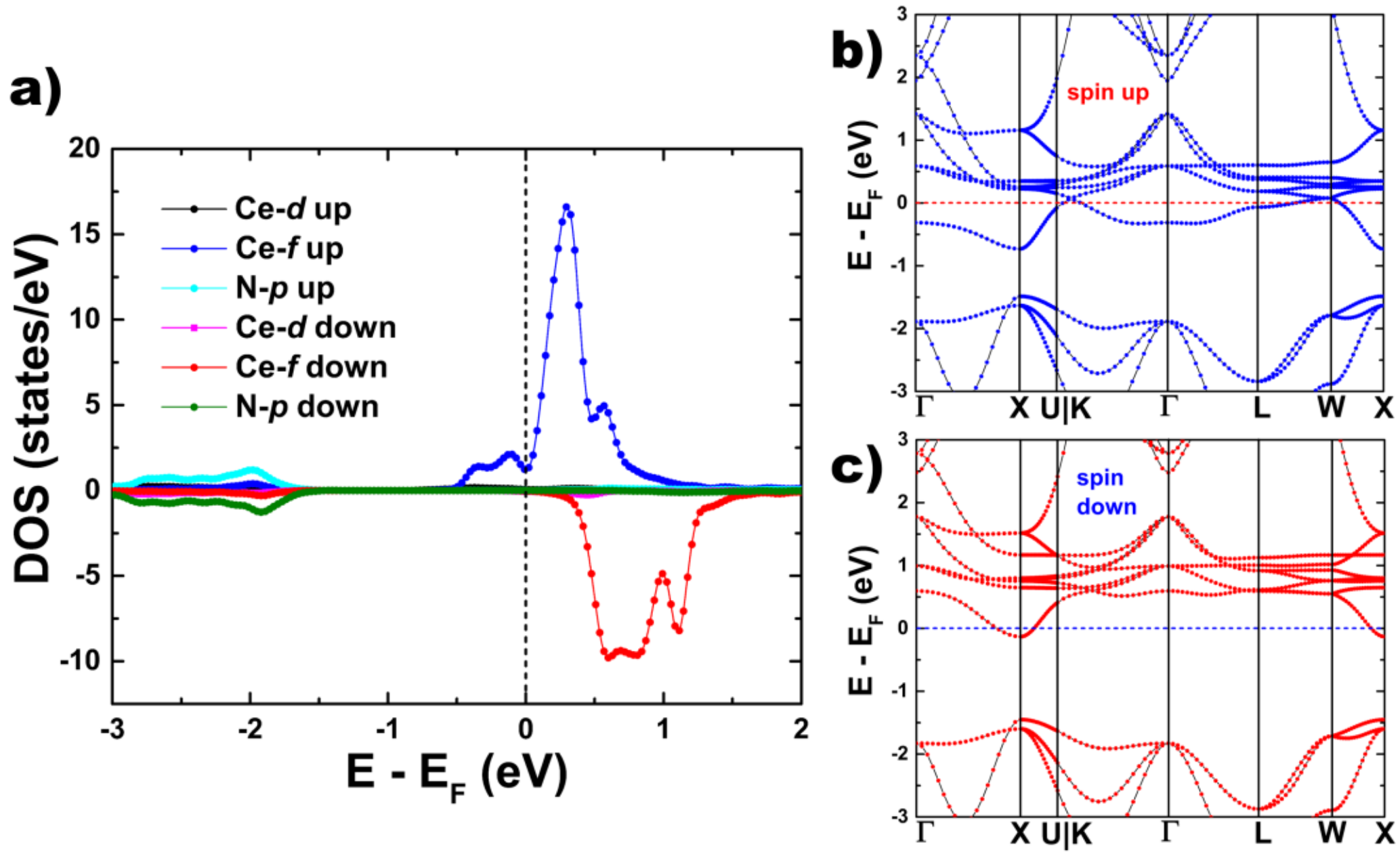

**Figure 2.** Electronic properties of bulk CeN, a) Projected density of states, b) spin up band structure and c) spin down band structure.

From figure 2-a, we can observe that the main contribution to electronic states is coming from Ce-*f* orbitals at energy values around the Fermi level, a few contributions are due to Ne-*p* orbitals for energies below the Fermi level. A quasi-half-metallic behavior can be inferred, as we observe a small crossing of one band around X-high-symmetry point very near Fermi-level, followed by a wide gap below Fermi level in the spin down channel. A metallic behavior can be seen for spin-up channel. The bands crossing the Fermi level are formed for these orbitals for both spin channels. In summary, bulk FM CeN is therefore characterized as a spin-polarized metal rather than a semiconductor or a fully half-metal.

In figure 3 we show the spin-resolved band structures and projected density of states (PDOS) for the 2D FM monolayer. We can appreciate a change in properties of spin-up channel, a small gap of 0.08 eV is opened (the highest occupied and lowest unoccupied bands approach $E_F$ very closely) in contrast to bulk counterpart, where a full metallic behavior is observed. From results, we can conclude that the FM monolayer can therefore be described as a narrow-gap or nearly gapless spin-polarized system.

On the other hand, for spin-down channel, the same behavior of bulk is preserved. The PDOS exhibits a pronounced spin asymmetry, particularly for the Ce-4$f$ states, and identifies these localized states as the main source of magnetic polarization in the FM phase. An occupied-states contribution is located approximately from 0 to 3 eV below the Fermi, with major contribution coming from Ce-$f$ and N-$p$ orbitals, followed by a depletion of states just above Femi level, and finally a region of high density of states due to Ce-f orbitals, at level energies between 1 and 3.5 eV for both spin channels, although with markedly different spectral distributions.

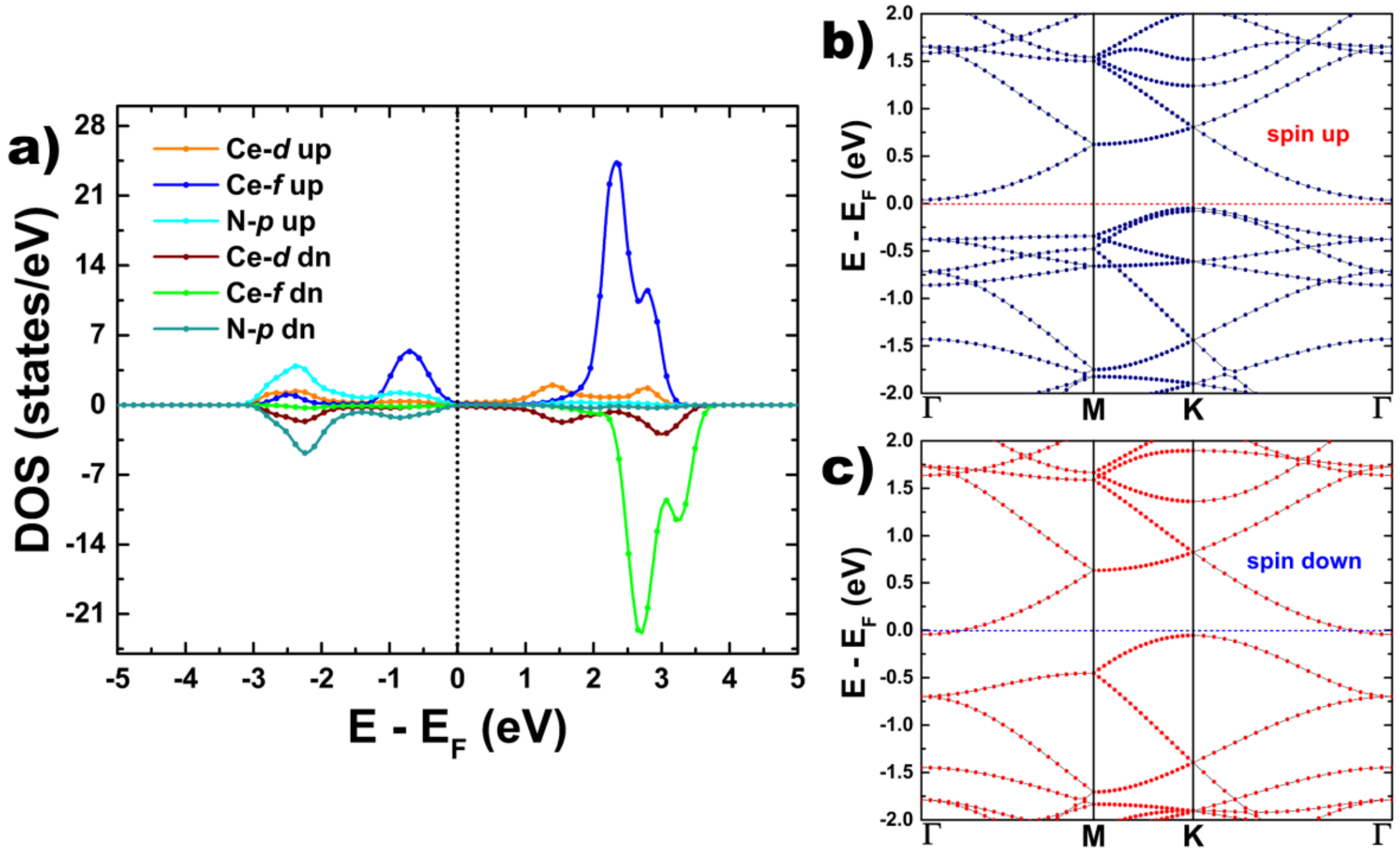

**Figure 3.** Electronic properties of FM monolayer, a) projected density of states, b) band structure of spin up channel and c) band structure of spin down channel.

As observed in bulk, the main contribution to states comes from Ce-$f$ orbitals, especially for energy values around Fermi level. However, the distribution of electronic states changes with respect to bulk. For spin up channel, the edge of last valence band is formed by Ce-$f$ orbitals, but a small contribution of N-$p$ orbitals is also observed. On the other hand, with respect to spin down channel, the band crossing the Fermi level is mainly formed by N-$p$ and Ce-$d$ orbitals, and the band just below the Fermi level is formed essentially by N-$p$ orbitals. This new behavior is attributed to the reduction in dimensionality, in terms of the interaction between N-$p_z$ orbitals and Ce-$f$ orbitals. A simultaneous presence of N-$p$ and Ce-$d$ orbitals over an energy range below Fermi level indicates appreciable hybridization between Ce and N orbitals. In general, the N-$p$ and Ce-$d$ contributions extend over broader energy intervals, which is consistent with their more itinerant character. This distinction is also reflected in the band structure, where relatively weakly dispersive states appear in the energy range associated with the Ce-*f* contribution, whereas more strongly dispersive bands are observed in regions dominated by N-*p* and Ce-*d* states.

With respect to AFM monolayer, the results are shown in figure 4. In this configuration, the highest occupied and lowest unoccupied bands approach Fermi level very closely, leading to the generation of a small indirect band gap of 0.08 eV in both spin channels. The valence-band maximum is located at K high-symmetry point, whereas the conduction-band minimum occurs at Γ point, indicating a small indirect K – Γ band gap. Consequently, changing the magnetic configuration from FM to AFM primarily modifies the spin organization of the Ce local moments, while producing comparatively smaller changes in the overall dispersion of the bands close to the Fermi energy, as the AFM phase displays a qualitatively similar orbital distribution to that of FM phase, this suggests that the electronic states defining the band edges are less strongly affected by the magnetic ordering than the localized Ce-*f* states responsible for the local moments.

With respect to the orbital distribution, the deeper valence states again contain substantial N-*p* and Ce-*f* contributions, followed by a depletion region of states just above from Fermi level up to 1 eV, and finally, a region of localized Ce-f states from 1 to 3.5 eV dominating the part of the unoccupied spectrum. This suggests that the electronic states defining the band edges are less strongly affected by the magnetic ordering than the localized Ce-$4f$ states responsible for the local moments. The edge of last valence band is mainly formed by Ce-*f* orbitals, with some contribution of N-*p* orbitals. On the other hand, the edge of first conduction band is originated by the contribution of Ce-*d* orbitals. The most significant difference with respect to the FM phase is found in the spin character of the electronic bands. As shown in Fig. 4, the spin-up and spin-down bands of the AFM configuration are symmetric throughout the whole *k*-path. This spin degeneracy is consistent with a compensated AFM state in which symmetry-related Ce atoms carry oppositely oriented local magnetic moments. Thus, the absence of net magnetization does not imply the disappearance of local Ce magnetism. Instead, finite Ce moments remain present but compensate through their antiparallel arrangement.

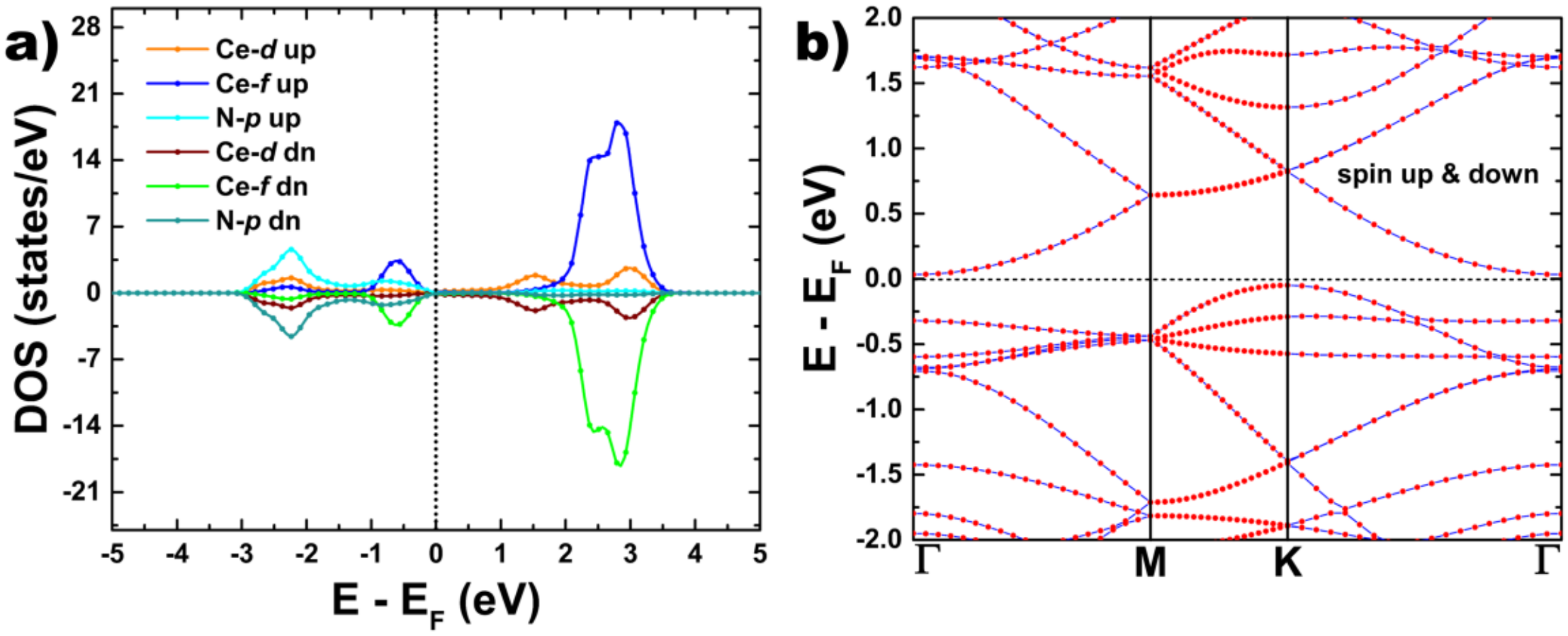

**Figure 4.** Electronic properties of AFM monolayer, a) projected density of states and b) band structure of spin up and down channels.

From results, we can conclude that the calculations reveal a marked reconstruction of the electronic states near the Fermi level upon going from the three-dimensional bulk structure to the 2D system, while the Ce-*f* orbitals retain a central role in the magnetic properties. On the other hand, a strong depletion of electronic states around Fermi level is observed in both phases of 2D system, in contrast to the behavior of bulk counterpart. The bulk PDOS shows strong Ce-*f* spectral weight much closer to the Fermi level than in the two-dimensional system. In particular, the spin up Ce-*f* contribution develops an intense structure immediately above Fermi level, whereas the spin down Ce-*f* contribution is predominantly distributed at somewhat higher energies. The difference between the two spin channels again demonstrates substantial exchange splitting and confirms the central role of Ce-*f* states in the magnetism of bulk CeN. However, their energetic distribution near to Fermi level differs markedly from that obtained in the 2D system, for both magnetic configurations.

The comparison between bulk and 2D CeN consequently reveals a pronounced dimensionality-induced reconstruction of the low-energy electronic structure. In the bulk system, Ce-derived bands contribute directly at the Fermi level, resulting in metallic behavior. In the 2D phase, by contrast, the density of states around Fermi level is strongly suppressed and no clear bands crossing is observed, resulting in a narrow-gap or nearly gapless electronic structure. In addition, whereas the dominant unoccupied Ce-*f* orbitals in bulk CeN occur relatively close to Fermi level, the corresponding strong Ce-*f* states in the 2D system are located substantially farther above the Fermi energy, suggesting a substantial reorganization of the Ce-*f* orbitals because of reduction in dimensionality.

The origin of this behavior can be associated with the modification of the local coordination environment and electronic hopping pathways produced by dimensional reduction. In bulk CeN, the three-dimensional crystal structure permits orbital overlap and electronic hopping along all spatial directions. Formation of the 2D structure removes the out-of-plane coupling and changes the coordination of the Ce and N atoms, thereby modifying Ce–N and Ce–Ce orbital overlap. Consequently, the hybridization between the Ce-*f*, Ce-*d*, and N-*p* states is altered, producing a reconstruction of the bands in the vicinity of the Fermi level. Reduced coordination may also favor a decrease in electronic bandwidth and an increased tendency toward localization. The observed separation of the dominant Ce-*f* spectral weight from the Fermi level region in the 2D structure is consistent with a modification of the balance between localized and itinerant behavior. The results are interpreted as evidence of modified Ce-*f* hybridization and a tendency toward enhanced localization induced by reduced dimensionality.

Despite the pronounced changes close to Fermi level, the deeper valence electronic structure retains some common characteristics in the bulk and 2D systems. In both cases, N-*p* orbitals hybridize with Ce-*f* orbitals, indicating that Ce–N hybridization remains an important component of the chemical bonding. Thus, dimensional reduction does not eliminate Ce–N hybridization but modifies its strength and its relationship with the Ce-*f* orbitals. This interplay between relatively localized Ce-*f* states and the more itinerant *d* and *p* derived states provides a consistent description of the electronic structure in both dimensionalities.

Magnetic ordering and dimensionality therefore affect different aspects of the electronic structure. A change from FM to AFM configurations in the 2D system primarily changes the relative orientation of the local Ce moments, leading from spin-split FM bands to an essentially spin-degenerate AFM electronic structure, while preserving the narrow-gap character. In contrast, dimensional reduction from bulk to the 2D structure produces a much stronger reconstruction of the states around the Fermi level, changing the system from a spin-polarized metallic bulk phase to a narrow-gap or nearly gapless two-dimensional state. Overall, the bulk-to-2D comparison demonstrates that reduced dimensionality provides an effective mechanism for tuning the electronic properties of CeN without suppressing its underlying Ce-*f* magnetism.

*3.4 Magnetic properties*

The magnetic behavior of CeN was investigated by comparing different spin configurations in the two-dimensional monolayer and bulk structures. Particular attention was paid to the stability and magnitude of the local Ce magnetic moments and to how these properties evolve upon dimensional reduction. The calculations reveal an important difference between bulk and two-dimensional CeN: whereas the monolayer supports well-defined local Ce moments in both ferromagnetic (FM) and antiferromagnetic (AFM) configurations, an AFM initialization in bulk CeN leads to a complete collapse of the local magnetic moments. This pronounced change indicates that dimensional reduction substantially modifies the magnetic energy landscape of CeN. For the CeN monolayer, FM, AFM, and nonmagnetic (NM) configurations were investigated. The calculated total energies are −57.291, −57.274, and −54.441 eV per computational cell for the FM, AFM, and NM states, respectively. The FM configuration therefore corresponds to the lowest-energy solution among the states considered.

The energy difference between the AFM and FM configurations is: $\Delta E_{\mathrm{AFM-FM}} = E_{\mathrm{AFM}} - E_{\mathrm{FM}} = 17.10\ \mathrm{meV/cell}$. Thus, although the FM configuration is energetically preferred, the AFM state is relatively close in energy. In contrast, the NM state lies approximately: $E_{\mathrm{NM}} - E_{\mathrm{FM}} \approx 2.85\ \mathrm{eV/cell}$ above the FM state. The large difference between the magnetic and NM solutions, compared with the much smaller FM–AFM separation, suggests the existence of two distinct energy scales. Formation of the local magnetic moments provides a large energetic stabilization, whereas the energy associated with changing their relative orientation from FM to AFM is considerably smaller. Consequently, the results indicate a strong tendency toward local-moment formation in the monolayer, while FM and AFM arrangements represent competing magnetic configurations once these moments are established.

The site-resolved magnetic moments reinforce this interpretation. In the FM monolayer, each Ce atom carries a moment of $m_{\mathrm{Ce}}^{\mathrm{FM}} = +0.921\ \mu_B$, whereas the N atoms exhibit a small antiparallel induced moment of $m_{\mathrm{N}}^{\mathrm{FM}} = -0.041\ \mu_B$.

In the AFM configuration, alternating Ce atoms carry moments of $m_{\mathrm{Ce}}^{\mathrm{AFM}} = \pm 0.919\ \mu_B$, while the corresponding N moments are much smaller, equal to $m_{\mathrm{N}}^{\mathrm{AFM}} = \pm 0.011\ \mu_B$.

A particularly important result is that the magnitude of the Ce local moment is essentially unchanged when the magnetic ordering is changed from FM to AFM: $|\ m_{\mathrm{Ce}}^{\mathrm{FM}}\ | = 0.921\ \mu_B$, and $|\ m_{\mathrm{Ce}}^{\mathrm{AFM}}\ | = 0.919\ \mu_B$. The difference is only about $0.002\ \mu_B$. Therefore, the transition between the calculated FM and AFM configurations primarily changes the relative orientation of the Ce moments rather than their magnitude. This behavior indicates that the Ce local moments are robust in the two-dimensional structure.

In both magnetic 2D configurations, approximately 98% of the site-projected Ce magnetic moment originates from the 4*f* contribution. The 5$d$ contribution is comparatively small, while the $s$ and $p$ contributions are negligible. The magnetic behavior of the CeN monolayer can consequently be identified as predominantly Ce-4*f* in character. The same occurs for the FM bulk.

Calculations also reveal a small spin polarization on the N atoms. In the FM monolayer, the N moment (associated with the N-*p* states) of approximately $-0.041\ \mu_B$ is oriented antiparallel to the Ce moments. This observation indicates that the magnetic response is not entirely confined to isolated Ce-*f* orbitals. Instead, the N states respond to the magnetic configuration of the neighboring Ce atoms, consistent with electronic coupling and hybridization between Ce- and N-derived states. The difference between the FM and AFM nitrogen moments is particularly informative. The magnitude decreases from approximately $0.041\ \mu_B$ in the FM state to approximately $0.011\ \mu_B$ in the AFM state. In the FM arrangement, the parallel Ce moments produce a coherent spin-polarized environment around the N atoms. In the AFM configuration, contributions associated with oppositely oriented neighboring Ce moments can partially compensate, resulting in a considerably weaker induced polarization on N. The sensitivity of the N polarization to the magnetic arrangement therefore provides additional evidence that Ce–N electronic interactions participate in the magnetic response of the monolayer.

The bulk FM state also exhibits pronounced local Ce magnetism. The calculated Ce moment is $m_{\text{Ce}}^{\text{bulk,FM}} = 0.973\ \mu_B$, the computed magnetic moment is in good agreement with the reported value of 0.90 μB **[29, 30]**. On the other hand, the N atoms exhibit a small antiparallel moment of $m_{\text{N}}^{\text{bulk,FM}} = -0.032\ \mu_B$. Comparison with the FM monolayer shows that the Ce moment decreases moderately from approximately $0.973\ \mu_B$ in bulk to $0.921\ \mu_B$ in the two-dimensional structure. Thus, dimensional reduction does not suppress the local Ce magnetism; the Ce moment remains close to $1\ \mu_B$.

A much more pronounced difference between bulk and monolayer CeN emerges when an AFM configuration is considered. When bulk CeN is initialized with an AFM arrangement, the self-consistent calculation does not preserve finite oppositely oriented Ce moments. Instead, the local moments collapse during electronic relaxation. The final site-projected magnetization is essentially zero on every Ce and N atom, including a vanishing Ce-*f* contribution. In other words, the converged bulk state should be regarded as a moment-collapsed solution obtained from an AFM initialization rather than as a stable AFM state. The moment-collapsed state therefore lies approximately 3.103 eV/cell above the FM state; the AFM initialization is unable to sustain a stable local-moment solution in bulk CeN.

The contrast with the monolayer is particularly striking. In bulk CeN, an AFM initialization results in the disappearance of the Ce local moments, whereas in the monolayer the AFM state preserves robust moments of approximately $\pm 0.919\ \mu_B$, dominated by the Ce-*f* orbitals. Furthermore, this well-defined AFM state lies only 17.1 meV/cell above the FM configuration. The calculations therefore indicate that dimensional reduction does more than slightly modify the magnitude of the Ce magnetic moment. It qualitatively changes the ability of the system to sustain an antiferromagnetically ordered local-moment solution. This behavior can be understood in terms of the substantial modification of the local electronic environment produced by dimensional reduction. Transforming bulk rocksalt CeN into a monolayer reduces the atomic coordination and changes the network of Ce–N interactions. Such changes are expected to modify orbital overlap, Ce–N hybridization, and the effective magnetic interactions connecting neighboring Ce sites. The reduced coordination of the two-dimensional structure provides a modified hybridization environment in which the Ce-*f* moments remain well defined even when neighboring moments are oppositely aligned. In contrast, within the three-dimensional bulk environment, the AFM initialization evolves toward a solution in which the local magnetic moments disappear. Within the monolayer, the modified electronic environment allows both FM and AFM local-moment states to exist, with very similar Ce-*f* moment magnitudes and a relatively small energy separation.

*3.5 Phonon dispersion and dynamical stability*

The lattice dynamics of the two-dimensional CeN monolayer were investigated for the ferromagnetic (FM) and antiferromagnetic (AFM) configurations by calculating their phonon dispersions along the high-symmetry path: Γ→M→K→Γ of the hexagonal Brillouin zone. Each spectrum contains six phonon branches: three acoustic and three optical modes. This number is consistent with the two-atom primitive cell used to construct the dynamical matrix. Results of phonon spectra of both configurations are depicted in figure 5.

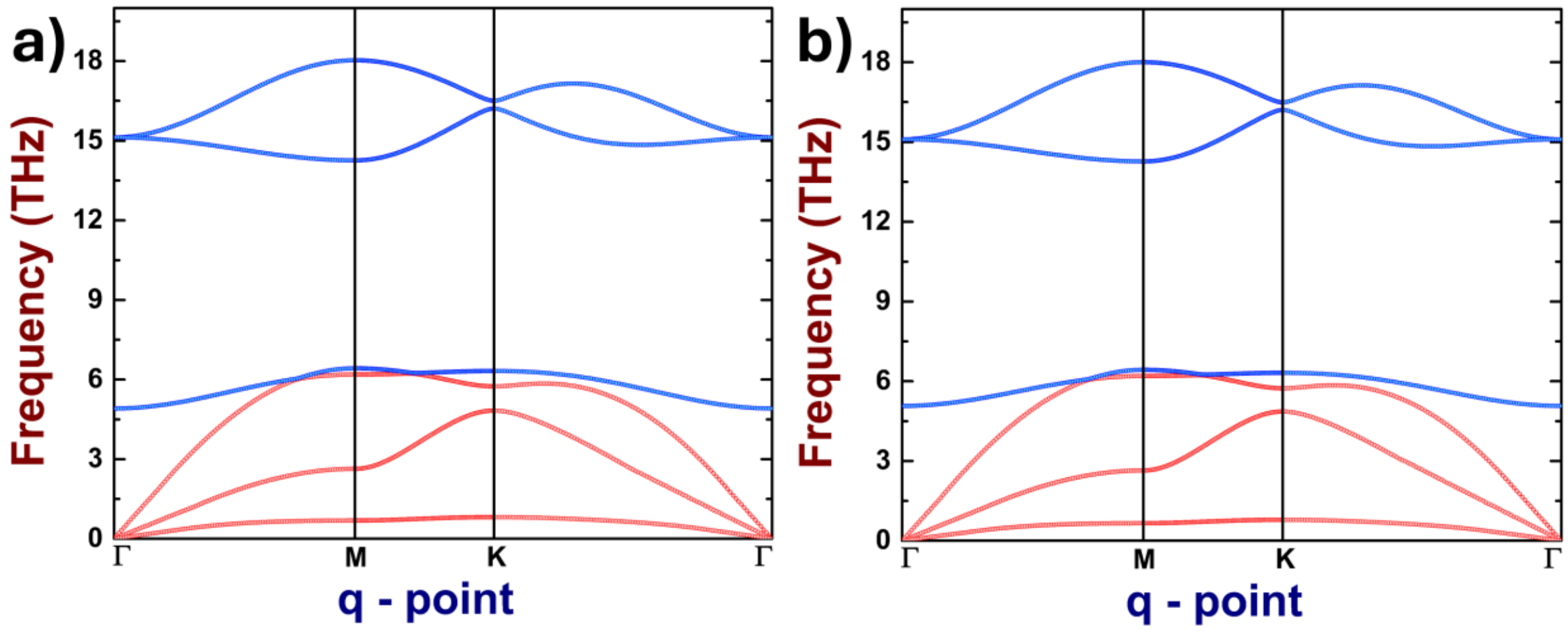

**Figure 5.** Phonon spectra of a) ferromagnetic 2D-CeN and b) antiferromagnetic 2D-CeN.

No imaginary phonon frequencies are observed in either magnetic configuration along the investigated path. This indicates that both FM and AFM configurations are dynamically stable along the sampled high-symmetry directions within the harmonic approximation; no dynamical instability is found along Γ→M→K→Γ path.

At the Γ point, the three acoustic branches converge to zero frequency, this behavior is required by translational invariance. The branch with the largest slope near Γ can be tentatively associated with the longitudinal acoustic mode (LA), while the branch with an intermediate slope can be assigned to the in-plane transverse acoustic mode (TA). The lowest and comparatively flat branch is tentatively identified as the out-of-plane flexural acoustic mode (ZA). For a free-standing two-dimensional material, the in-plane acoustic modes show linear behavior near Γ. In contrast, the flexural acoustic branch follows a different dispersion. This explains why the ZA mode appears considerably flatter than the LA and TA branches. Physically, the low frequency of this mode reflects the relatively small energy required to bend a two-dimensional sheet compared with the energy required to stretch or shear its in-plane bonds.

The lowest acoustic branch remains positive over the complete path in both configurations. Its frequencies are 0.688 and 0.817 THz at M and K, respectively, for the FM phase and 0.660 and 0.789 THz for the AFM phase. Therefore, the low frequency and flat character of this branch should be interpreted as features of a flexural ZA vibration rather than as evidence of dynamical instability. The acoustic branches of the two magnetic phases are remarkably similar. This is consistent with their nearly identical elastic constants, because the slopes of the acoustic branches near Γ are related to the elastic stiffnesses and sound velocities. In particular, both phases have an in-plane Young's modulus of approximately 33.5 N/m. All details in regards the elastic constants will be provided in the next subsection.

The optical sector consists of one low-frequency branch and two high-frequency branches. The two high-frequency optical branches are degenerate at Γ in both magnetic configurations. This degeneracy is consistent with the in-plane symmetry of the hexagonal lattice. Away from Γ, their degeneracy is lifted, with the largest separation occurring near the M point. The lowest optical branch increases from 4.909 THz at Γ to 6.423 THz at M in the FM phase. In the AFM phase, it increases from 5.070 to 6.429 THz. The frequency range of this optical branch overlaps with that of the upper acoustic branches. Therefore, the system does not possess a complete gap between all acoustic and optical modes. Acoustic and optical branches have comparable symmetries and similar frequencies; they may hybridize or exchange polarization character. The maximum calculated frequency is approximately 18.032 THz for the FM phase and 18.001 THz for the AFM phase. The difference of only 0.031 THz shows that the overall phonon-frequency range is essentially unaffected by magnetic ordering.

Both phonon spectra exhibit a pronounced separation between the four low-frequency branches and the two upper optical branches. For the FM configuration, the highest frequency in the lower group is approximately 6.423 THz, while the lowest frequency in the upper group is approximately 14.259 THz. For the AFM configuration the corresponding values are equal to 6.429 THz and 14.269 THz respectively. Thus, both configurations exhibit essentially the same gap equal to 7.84 THz. This feature should be described as a phonon band gap between low- and high-frequency mode groups, rather than as a conventional acoustic–optical gap. The lowest optical branch lies below the gap and overlaps in frequency with the acoustic sector. The pronounced frequency separation is likely related to the large atomic-mass difference between Ce and N. The heavier Ce atoms are therefore expected to contribute mainly to the low-frequency modes, while the lighter N atoms are expected to dominate the high-frequency optical modes.

The phonon spectra of the two magnetic configurations are remarkably similar. Their acoustic branches, high-frequency optical branches, maximum frequencies, and phonon gaps differ only slightly. The most noticeable difference occurs in the lowest optical mode at Γ. Its frequency increases from 4.909 THz (FM) to 5.070 THz (AFM), the frequency shift is 0.161 THz (AFM -

FM), corresponding to a relative hardening of approximately 3.3%. This is the largest mode-dependent difference between the two phases. It indicates that this particular vibration is more sensitive to the magnetic arrangement than the remaining modes. Its displacement pattern may modify Ce–Ce distances, Ce–N–Ce bonding geometry, or magnetic exchange pathways more strongly.

Because the atomic masses are identical in the FM and AFM calculations, differences between their phonon frequencies can arise only from changes in the interatomic force constants. In physical terms, the Ce–N bonding network and its response to atomic displacements are only weakly modified when the relative orientation of the Ce magnetic moments changes. The lattice dynamics are therefore governed predominantly by chemical bonding rather than by magnetic ordering. In conclusion, magnetic ordering may strongly affect the spin-resolved electronic structure while causing only minor changes in the curvature of the total energy with respect to atomic displacement.

*3.5 Elastic properties and mechanical stability*

The elastic response of the 2D CeN monolayer was investigated for both ferromagnetic (FM) and antiferromagnetic (AFM) configurations. Because the calculations were performed using a periodic slab containing vacuum along the out-of-plane direction, the elastic constants obtained in GPa depend on the simulation-cell height and were consequently converted into two-dimensional stiffness coefficients in (N/m) units according to:

$$\mathrm{C_{ij}^{2D}[N/m] = 0.1L_z[\AA]C_{ij}^{3D}[GPa]} \qquad (4),$$

where $\mathrm{L_z}$ corresponds to vacuum level assigned to the supercell to avoid interaction between adjacent layers (the out-of-plane lattice parameter), equal to 15 Å in our calculations.

Once computing the elastic constants, it is possible to compute other mechanical properties such as the 2D Young's modulus, Poisson ratio, the in-plane shear modulus, and the two-dimensional layer modulus.

By considering the hexagonal symmetry of the system, we have just five independent elastic constants: $C_{11}$, $C_{12}$, $C_{13}$, $C_{33}$ and $C_{44}$. An extra elastic constant $C_{66}$ can be computed as:

$$C_{66} = \frac{C_{11} - C_{12}}{2} \qquad (5).$$

The other mechanical properties can be computed as follows:

The Young's modulus, that describes how strongly the monolayer resists a change in length when pulled within its plane:

$$Y_{2D} = \frac{(C_{11}^{2D})^2 - (C_{12}^{2D})^2}{(C_{11}^{2D})} \qquad (6).$$

The Poison's ratio, that indicates conventional lateral contraction: when the monolayer is stretched in one in-plane direction, it contracts in the perpendicular direction:

$$\nu_{2D} = \frac{C_{12}^{2D}}{C_{11}^{2D}} \qquad (7).$$

The in-plane shear modulus, which indicates a measure of the resistance to small in-plane shear deformations:

$$G_{2D} = C_{66}^{2D} \qquad (8).$$

The 2D layer modulus, that measures the resistance to uniform biaxial in-plane deformation:

$$K_{2D} = \frac{C_{11}^{2D} + C_{12}^{2D}}{2} \qquad (9).$$

The mechanical properties of 2D CeN structures for both configurations are summarized in table 2.

**Table 2.** Mechanical properties of FM and AFM 2D CeN.

| Magnetic state | $C_{11}^{2D}$ [N/m] | $C_{12}^{2D}$ [N/m] | $C_{13}^{2D}$ [N/m] | $C_{33}^{2D}$ [N/m] | $C_{44}^{2D}$ [N/m] | $C_{66}^{2D}$ [N/m] | $Y_{2D}$ [N/m] | $K_{2D}$ [N/m] | $\nu_{2D}$ |
|---|---|---|---|---|---|---|---|---|---|
| **AFM** | 59.58 | 39.57 | 0.09 | 1.01 | 1.81 | 9.82 | 33.52 | 49.58 | 0.668 |
| **FM** | 59.96 | 39.84 | 0.22 | 1.03 | 0.97 | 10.06 | 33.49 | 49.91 | 0.665 |

To assess the dynamical stability for 2D hexagonal systems, the necessary and sufficient conditions that must be fulfilled are **[31]**:

$$C_{11} > |C_{12}|, \quad 2(C_{13})^2 < C_{33}(C_{11} + C_{12}), \quad C_{44} > 0, \quad C_{66} > 0 \qquad (10).$$

By considering the results of table 2 we can conclude that both magnetic phases of 2D CeN are mechanically stable, as the four conditions are fulfilled. This elastic stability is consistent with the phonon calculations, which show no imaginary frequencies along the investigated $\Gamma - M - K - \Gamma$ path for either magnetic phase.

The calculated out-of-plane components, such as $C_{33}$ and $C_{44}$ are considerably smaller than the in-plane components. These quantities should not, however, be interpreted in the same manner as the intrinsic in-plane stiffnesses. These components depend strongly on the selected vacuum thickness and do not generally represent intrinsic elastic properties of an isolated monolayer, so the mechanical discussion should therefore focus primarily on the in-plane components.

On the other hand, the Young's moduli of the FM and AFM configurations are therefore practically identical. This indicates that changing the magnetic ordering has a negligible effect on the resistance of the monolayer to small in-plane uniaxial deformations. The positive Poisson ratios indicate conventional lateral contraction: when the monolayer is stretched in one in-plane direction, it contracts in the perpendicular direction. The relatively large value of approximately 0.66 indicates strong coupling between the two normal in-plane strain components and considerable transverse contraction under uniaxial tension. The FM shear modulus is approximately 2.4% larger than the AFM value. Thus, the FM configuration is only marginally more resistant to small in-plane shear deformations. The magnitude of this difference is too small to indicate a strong magnetically induced modification of the mechanical response. In regards the two-dimensional layer modulus, the difference between the two values is less than 1%, demonstrating that the two magnetic configurations possess essentially the same resistance to uniform biaxial deformation.

The close agreement between the FM and AFM values indicates that the strength and curvature of the Ce–N bonding potential are only weakly affected by the relative orientation of the Ce magnetic moments. In other words, the linear elastic response is governed predominantly by the chemical bonding network rather than by the magnetic arrangement. Overall, the limited differences between the two configurations suggest weak magnetoelastic coupling within the small-strain regime considered in the present calculations.

*3.6 Thermal Stability from Ab Initio Molecular Dynamics*

The finite-temperature stability of the CeN monolayer was investigated by means of ab initio molecular dynamics (AIMD) simulations for both the ferromagnetic (FM) and antiferromagnetic (AFM) configurations. These calculations provide a complementary assessment of the phonon and elastic calculations because they allow the structural response of the monolayer to thermal atomic motion to be examined directly. The AIMD simulations were performed at a target temperature of 300 K using spin-polarized DFT+U. A total of 1000 ionic steps were performed using a time step of 5 fs, resulting in a total simulation time of 5 ps.

The instantaneous temperature exhibits pronounced oscillations during both the FM and AFM simulations during the 1000 ionic steps. Despite these oscillations, the average temperature is very close to the imposed target value of 300 K; for the FM configuration, the average temperature calculated over the complete 1000-step trajectory is approximately 297.97 K. A very similar behavior is observed for the AFM configuration, in this case, the average temperature is approximately 295.90 K. The average temperatures of the two trajectories are therefore remarkably similar and differ by only about 2 K. This indicates that the FM and AFM structures were sampled under essentially equivalent thermal conditions, allowing their finite-temperature structural responses to be compared directly.

The energetic evolution shows a similarly close correspondence between the two magnetic configurations. For the FM trajectory, the average value of the monitored energy is approximately −55.92417 eV, with a standard deviation of 0.09352 eV. For the AFM trajectory, the corresponding average energy is approximately −55.92195 eV, with a standard deviation of 0.11219 eV. No catastrophic energetic divergence is observed for either magnetic configuration. The absence of an energetic runaway in either magnetic configuration is consistent with the persistence of the atomic structures discussed below. The results are shown in figure 6.

The very similar temperature and energy fluctuations observed for the FM and AFM configurations indicate that both systems exhibit comparable finite-temperature behavior under the conditions considered here. Importantly, neither trajectory displays a progressive thermal or energetic divergence. This behavior, together with the preservation of the CeN framework observed in the initial and final structural configurations, depicted in figure 7, supports the finite-temperature structural stability of both magnetic states at 300 K over the investigated 5 ps timescale.

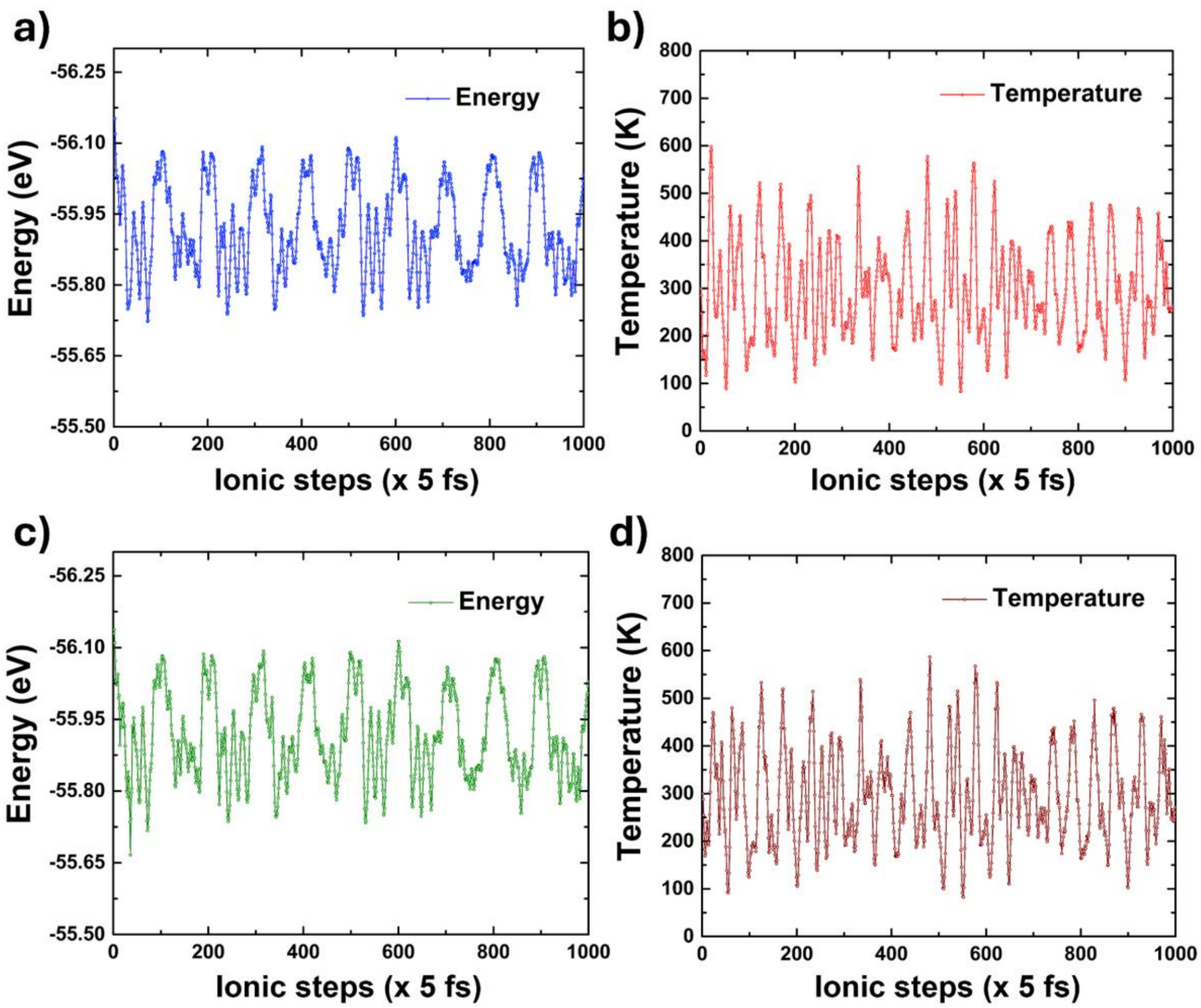

**Figure 6.** Ab initio molecular dynamics simulations, a) energy fluctuations for FM phase, b) temperature fluctuations for FM phase, c) energy fluctuations for AFM phase and d) temperature fluctuations for AFM phase.

Let us discuss the initial and final structures obtained from the MD simulations for the FM and AFM configurations depicted in Fig. 7. These structures provide a direct visualization of the response of the CeN monolayer to thermal excitation. At the beginning, the structures of both magnetic configurations are essentially planar, at the end of 5 ps, both structures show noticeable out-of-plane atomic displacements. However, comparison of the initial and final configurations clearly indicates that these distortions correspond primarily to thermal corrugation of the monolayer rather than to destruction of the CeN lattice. In both cases, the characteristic two-dimensional arrangement of the Ce and N atoms remains identifiable in the final structure.

For the FM configuration, the largest upward displacement relative to the initial structure is approximately +0.18 Å, while the largest downward displacement is approximately −0.21 Å. Consequently, the total vertical spread of the final FM structure is approximately 0.38 Å. The corresponding root-mean-square out-of-plane displacement is approximately 0.12 Å. The initially planar FM monolayer develops a moderate degree of corrugation after the AIMD simulation. Nevertheless, the Ce and N atoms remain associated with the original two-dimensional framework, with no obvious indication of fragmentation, major atomic migration, or catastrophic reconstruction. The observed deformation can therefore be interpreted as thermally induced rippling of the monolayer.

A similar behavior is observed for the AFM configuration. The largest upward displacement is approximately +0.24 Å, whereas the largest downward displacement is approximately −0.31 Å. The final AFM structure has a vertical spread of approximately 0.55 Å and a root-mean-square out-of-plane displacement of approximately 0.16 Å. Thus, the final AFM structure displays somewhat stronger corrugation than the corresponding FM structure. Nevertheless, the AFM monolayer retains the characteristic CeN framework and does not show evidence of structural collapse.

The most relevant result from the structural comparison is that both the FM and AFM monolayers preserve their characteristic two-dimensional CeN framework after 5 ps at 300 K. The loss of perfect planarity during AIMD is not necessarily an indication of structural instability. At finite temperature, the atoms vibrate around their equilibrium positions, and in a two-dimensional material these vibrations include an important out-of-plane component, therefore, an initially planar monolayer is expected to develop some degree of thermal rippling at finite temperature. The relevant distinction is between a moderate reversible distortion of the layer and a genuine structural transformation involving bond breaking, atomic migration, fragmentation, or reconstruction. The present simulations show the former behavior rather than the latter.

This interpretation is also qualitatively consistent with the phonon calculations. The phonon spectra of both magnetic configurations contain a relatively flat low-frequency acoustic branch. In a two-dimensional material, a low-frequency out-of-plane acoustic mode can facilitate thermally activated flexural motion. The moderate corrugation observed during MD ionic steps is therefore compatible with the vibrational characteristics obtained independently from the phonon calculations.

In general, the MD simulations reveal an overall similar structural response for the FM and AFM configurations. This result is consistent with the similarity observed between the FM and AFM phonon dispersions. Taken together, the phonon and AIMD results indicate that changing the relative orientation of the Ce magnetic moments does not produce a dramatic modification of the lattice response, in other words, there is a modest influence of magnetic ordering on the lattice dynamics.

Overall, the AIMD simulations support the finite-temperature structural stability of both magnetic configurations of the CeN monolayer at 300 K over the investigated 5 ps timescale. Taken together, the elastic constants, phonon spectra, and MD simulations provide complementary evidence for the structural robustness of the CeN monolayer. In particular, the preservation of the two-dimensional framework in both the FM and AFM final MD structures indicates that the monolayer can accommodate room-temperature thermal fluctuations without undergoing an obvious structural transformation.

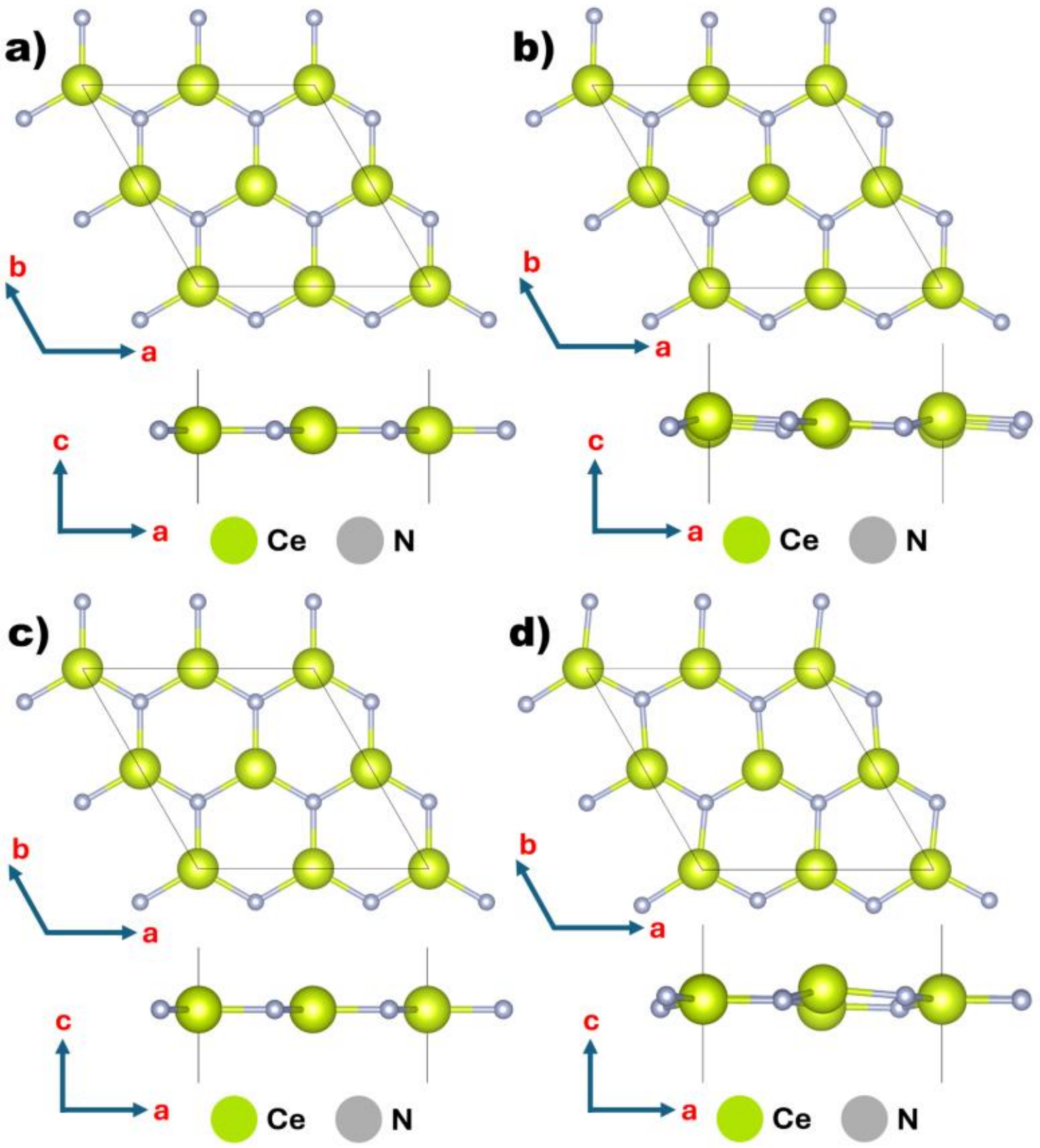

**Figure 7.** Structures at initial and final steps of MD simulations, a) top and side views of initial structure of FM configuration, b) top and side views of final structure of FM configuration, c) top and side views of initial structure of AFM configuration and d) top and side views of final structure of AFM configuration.

## 4. Conclusions

In this work, the structural, energetic, magnetic, electronic, mechanical, dynamical, and thermal properties of a novel two-dimensional CeN monolayer derived from the (111) orientation of the bulk rocksalt CeN structure were systematically investigated using first-principles calculations. The reduction from the three-dimensional bulk structure to the two-dimensional limit produces important changes in the CeN physical properties. In particular, the Ce–N coordination decreases from sixfold to threefold, accompanied by a significant shortening of the Ce–N bonds and the formation of an essentially planar hexagonal lattice. Dimensional reduction also modifies the electronic and magnetic behavior, allowing well-defined FM and AFM configurations with magnetic moments predominantly localized on the Ce atoms and mainly associated with the Ce 4*f* states. The FM configuration is slightly lower in energy than the AFM state, indicating a close energetic competition between different magnetic arrangements in the monolayer.

Interestingly, despite the clear influence of magnetic ordering on the electronic and magnetic properties, the underlying structural and mechanical characteristics are only weakly affected by the orientation of the magnetic moments. The FM and AFM configurations exhibit nearly identical equilibrium geometries, Ce–N bond lengths, elastic responses, and phonon spectra, indicating that these properties are governed predominantly by the Ce–N bonding network rather than by the specific magnetic arrangement. The elastic, phonon, and ab initio molecular dynamics results further support the mechanical, dynamical, and finite-temperature stability of the two-dimensional structure. Overall, our results show that dimensional reduction has a much stronger influence on the physical properties of CeN than the change between FM and AFM magnetic ordering, while the structural and mechanical framework remain remarkably robust against changes in the magnetic configuration. The coexistence of structural stability, localized Ce 4*f* magnetism, and spin-dependent electronic properties make the two-dimensional CeN as a novel interesting system for further investigations of magnetic 2D materials and potential spin-dependent and spintronic applications.

## Acknowledgments

JMGH acknowledges SECIHTI for the postdoctoral scholarship. Part of calculations were performed in the LNS supercomputer (BUAP). We also acknowledge IFUAP for providing the computational resources of Tlahuicole cluster for computing calculations.